%% file: main.tex
\documentclass[10pt,journal,compsoc]{IEEEtran}
\usepackage{tabu}                      
\usepackage{booktabs}                 
\usepackage{lipsum}                    
\usepackage{mwe}        
\usepackage{amsmath}
\usepackage{mathptmx}
\usepackage{bm}
\usepackage{makecell}
\usepackage{pifont} 
\usepackage{url}
\usepackage{color}

\DeclareMathAlphabet{\mathcal}{OMS}{cmsy}{m}{n}
\DeclareSymbolFont{largesymbols}{OMX}{cmex}{m}{n}

\newcommand{\rv}[1]{{\color{black}{#1}}}
\newcommand{\rrv}[1]{{\color{black}{#1}}}
\newcommand{\rvv}[1]{{\color{black}{#1}}}

\graphicspath{{figs/}{figures/}{pictures/}{images/}{./}}

\begin{document}

\title{Chart2Vec: A Universal Embedding of Context-Aware Visualizations}

\author{
Qing Chen, Ying Chen, Ruishi Zou, Wei Shuai, Yi Guo, Jiazhe Wang, and Nan Cao

\IEEEcompsocitemizethanks{\IEEEcompsocthanksitem Qing Chen, Ying Chen, Ruishi Zou, Wei Shuai, Yi Guo, and Nan Cao are with Intelligent Big Data Visualization Lab at Tongji University. Nan Cao is the corresponding author. \protect
E-mails: \{qingchen,chenying2929,zouruishi,shuaiwei,nan.cao\}@tongji.edu.cn
\IEEEcompsocthanksitem Jiazhe Wang is with Ant Group. Email: jiazhe.wjz@antgroup.com

}
}

\markboth{Journal of \LaTeX\ Class Files,~Vol.~X, No.~X, March~2024}%
{Shell \MakeLowercase{\textit{et al.}}: A Sample Article Using IEEEtran.cls for IEEE Journals}


\IEEEtitleabstractindextext{
\begin{abstract}
The advances in AI-enabled techniques have accelerated the creation and automation of visualizations in the past decade. However, presenting visualizations in a descriptive and generative format remains a challenge. Moreover, current visualization embedding methods focus on standalone visualizations, neglecting the importance of contextual information for multi-view visualizations. To address this issue, we propose a new representation model, Chart2Vec, to learn a universal embedding of visualizations with context-aware information. Chart2Vec aims to support a wide range of downstream visualization tasks such as recommendation and storytelling. Our model considers both structural and semantic information of visualizations in declarative specifications. To enhance the context-aware capability, Chart2Vec employs multi-task learning on both supervised and unsupervised tasks concerning the cooccurrence of visualizations. We evaluate our method through an ablation study, a user study, and a quantitative comparison. The results verified the consistency of our embedding method with human cognition and showed its advantages over existing methods.
\end{abstract}

\begin{IEEEkeywords}
Representation Learning, Multi-view Visualization, Visual Storytelling, Visualization Embedding
\end{IEEEkeywords}
}

\maketitle

\input{sections/1introduction}
\input{sections/2relatedwork}

\input{sections/3dataset}
\input{sections/4model}
\input{sections/5evaluation}
\input{sections/6discussion}
\input{sections/7conclusion}
\input{sections/acknowledgement}
\bibliographystyle{IEEEtran}
\bibliography{main}


\begin{IEEEbiography}[{\includegraphics[width=1in,height=1.25in,clip,keepaspectratio]{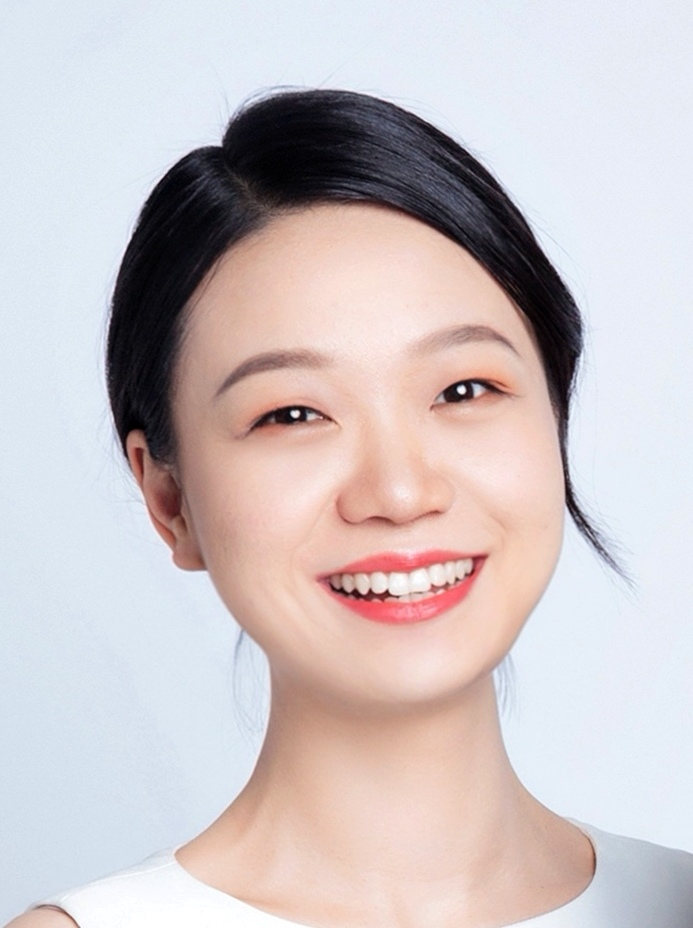}}]{Qing Chen} received her B.Eng degree from the Department of Computer Science, Zhejiang University and her Ph.D. degree from the Department of Computer Science and Engineering, Hong Kong University of Science and Technology (HKUST). After receiving her PhD degree, she worked as a postdoc at Inria and Ecole Polytechnique. She is currently an associate professor at Tongji University. Her research interests include information visualization, visual analytics, human-computer interaction, generative AI and their applications in education, healthcare, design, and business intelligence.
\end{IEEEbiography}

\begin{IEEEbiography}[{\includegraphics[width=1in,height=1.25in,clip,keepaspectratio]{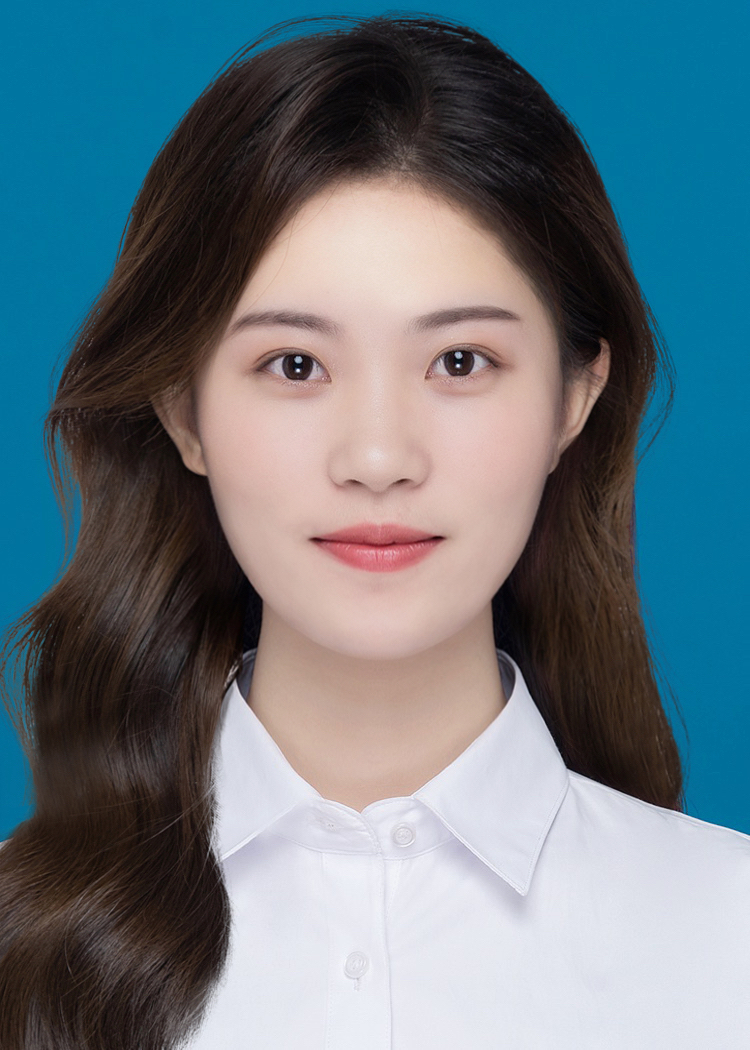}}]{Ying Chen} received her bachelor's degree from the Department of Artificial Intelligence and Computer Science, Jiangnan University in 2021. Currently, she is a master's candidate at Tongji University. Her research interests include data visualization, human-computer interaction, and the integration of artificial intelligence with business intelligence.
\end{IEEEbiography}

\begin{IEEEbiography}[{\includegraphics[width=1in,height=1.25in,clip,keepaspectratio]{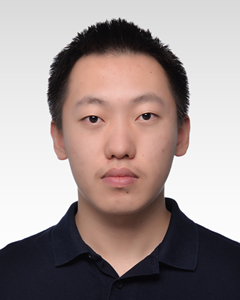}}]{Ruishi Zou} is pursuing his undergraduate degree from the Department of Computer Science at Tongji University. He is also a part of the Intelligent Big Data Visualization (iDV$^{x}$) Lab at Tongji University. His research interests include information visualization, human-AI interaction, and user interface software and technology.
\end{IEEEbiography}

\begin{IEEEbiography}[{\includegraphics[width=1in,height=1.25in,clip,keepaspectratio]{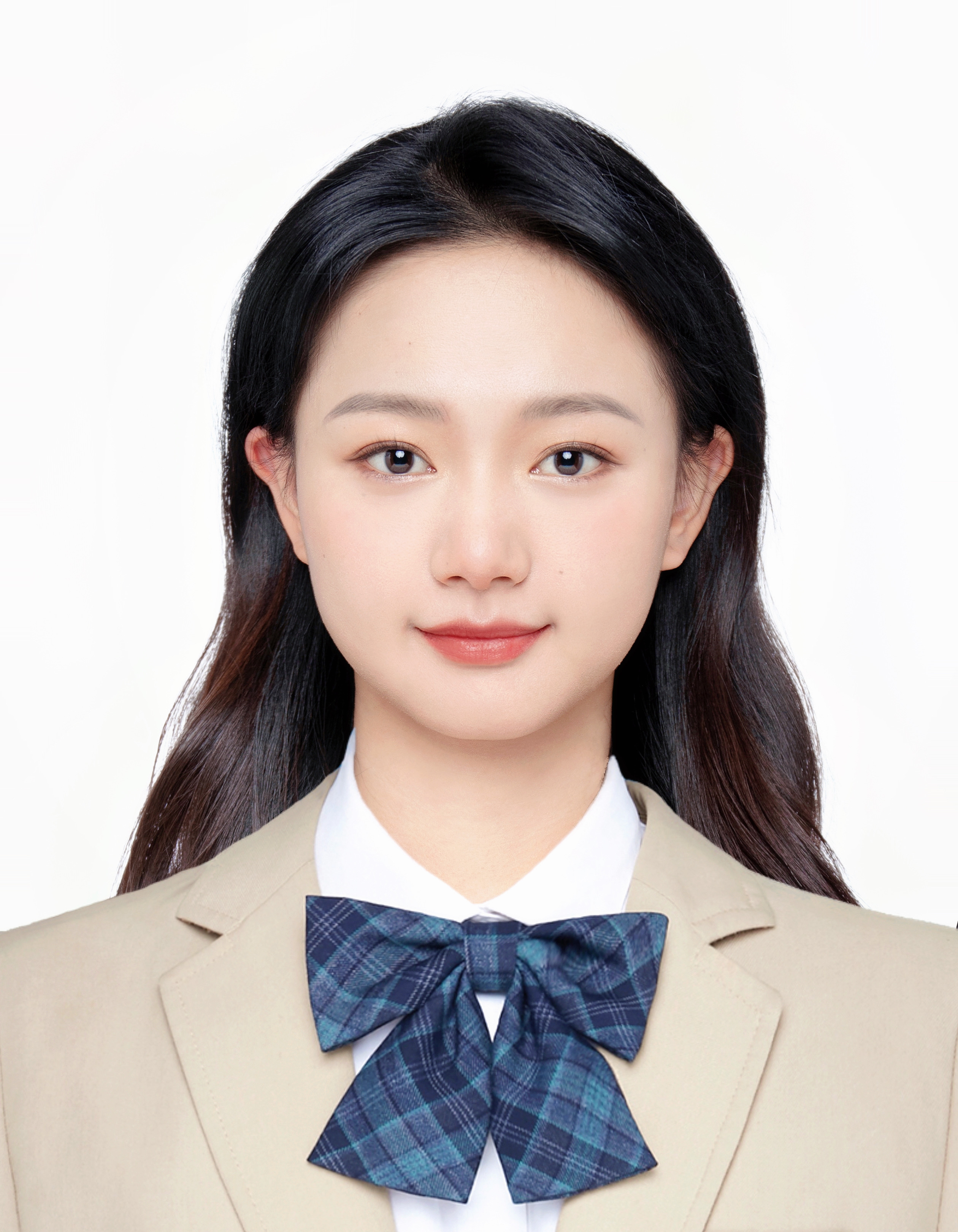}}]{Wei Shuai} received her bachelor's degree from the Department of Artificial Intelligence and Computer Science, Jiangnan University in 2022. Currently, she is a master's candidate at Tongji University. Her research interests include AI-supported design and information visualization.
\end{IEEEbiography}

\begin{IEEEbiography}[{\includegraphics[width=1in,height=1.25in,clip,keepaspectratio]{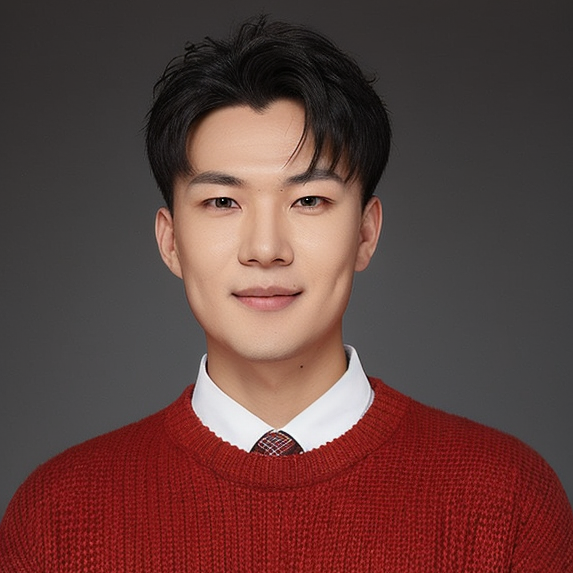}}]{Yi Guo} received his M.S. degree in Financial Mathematics from the University of New South Wales, Australia in 2019. He is currently working toward his Ph.D. degree as part of the Intelligent Big Data Visualization (iDV$^{x}$) Lab, Tongji University. His research interests include data visualization and deep learning.
\end{IEEEbiography}

\begin{IEEEbiography}[{\includegraphics[width=1in,height=1.25in,clip,keepaspectratio]{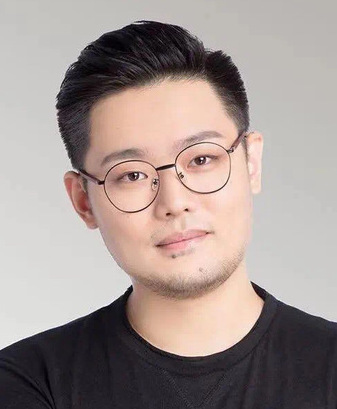}}]{Jiazhe Wang}  holds a Master's degree in Computer Science from the University of Oxford and is currently pursuing a part-time Ph.D. at Tongji University. He serves as the tech lead of the Intelligent Agent Platform at Alibaba Group. Previously, he was a pivotal member of AntV, Ant Group's data visualization team, and as a tech leader in augmented analytics at Ant Group. His research interests are primarily in artificial intelligence, agent technologies, visual analytics, and augmented analytics.
\end{IEEEbiography}

\begin{IEEEbiography}[{\includegraphics[width=1in,height=1.25in,clip,keepaspectratio]{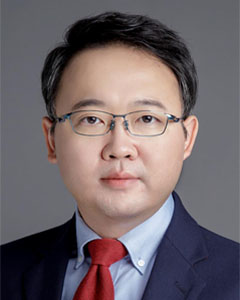}}]{Nan Cao} received his Ph.D. degree in Computer Science and Engineering from the Hong Kong University of Science and Technology (HKUST), Hong Kong, China in 2012. He is currently a professor at Tongji University and the Assistant Dean of the Tongji College of Design and Innovation. He also directs the Tongji Intelligent Big Data Visualization Lab (iDV$^x$ Lab) and conducts interdisciplinary research across multiple fields, including data visualization, human computer interaction, machine learning, and data mining. He was a research staff member at the IBM T.J. Watson Research Center, New York, NY, USA before joining the Tongji faculty in 2016.
\end{IEEEbiography}

\end{document}

%% file: sections/1introduction.tex
\section{Introduction}
Data visualizations are an important means to help people quickly identify complex data patterns and communicate insights.
\rrv{Automatic methods help accelerate the visualization creation process by improving the quality of the dataset used for visualization~\cite{wongsuphasawat2016voyager}, extracting the most meaningful information or insights from the data~\cite{shi2020calliope,cui2019datasite}, and selecting the appropriate visual representation~\cite{hu2019vizml,draco2019}. This allows users to grasp the key information from visualizations more quickly, accurately, and comprehensively~\cite{luo2018deepeye}.}
With the abundance of visualizations created by experts and automated systems, visualization themselves have become a new format of data~\cite{wu2021ai4vis}. Therefore, it is worth studying how to effectively and precisely represent such visualization data in a generalizable format to support downstream visualization and visual analytics tasks such as comparison~\cite{chen2020composition}, recommendation~\cite{oppermann2020vizcommender}, assessment~\cite{fu2019visualization}, and querying~\cite{ma2018scatternet}.

Inspired by the recent advances in representation learning, several studies in the visualization community attempted to use semantic vectors (i.e., embeddings~\cite{demiralp2014visual}) to represent information from the visualization data. For example, ChartSeer~\cite{zhao2020chartseer} adopted an encoder-decoder approach that converts visualization to and from embeddings to assist exploratory visual analysis. More recently, Li et al.~\cite{li2022structure} proposed a structure-aware method to improve the performance of visualization retrieval by collectively considering both visual and structural information. Compared to heuristic or rule-based methods, representation learning allows a more flexible and general presentation of visualization. Once the embedding features have been learned, they can be applied to a variety of downstream visualization tasks. 

Nevertheless, these attempts to present visualizations through embeddings can only be applied to one or two specific visualization tasks such as visual comparison and visualization recommendation. Moreover, most existing work is focused on visualization tasks for a single-view visualization. When considering multi-view visualizations, context information is a critical aspect of visualization representation that influences the outcome of subsequent tasks. 
There is still a lack of a universal representation of visualizations that can be used for various downstream tasks while taking contextual information into account.


To fill this research gap, we aim to propose a universal embedding of context-aware visualizations based on the associations derived from a large corpus of \rrv{visualizations specifications}. \rrv{In this paper, context-aware refers to the co-occurrence and logical relationships within multi-view visualizations, such as narrative sequences in data stories and logical orders of visual analytic findings.} The remaining challenges are as follows. 
First, we need to formulate a proper input embedding that leverages both the semantic content and the structural information of each visualization. To achieve this, we reviewed related studies on natural language descriptions of visualization data~\cite{lundgard2021accessible, stokes2023striking}, then summarized the key structural features that can be obtained from \rrv{visualizations specifications}. Compared to existing methods which only extract explicit information, such as chart types and data statistics (sometimes referred to as ``data facts''~\cite{srinivasan2018augmenting}), our input chart embedding of the proposed model also considers implicit information,
\rrv{ including specific field-related details from the dataset, such as field names, corresponding values, and data semantics.} Second, a large-scale dataset of context-aware visualizations is required to form the training and test datasets. Multiple visualizations in a cohesive representation can be regarded as multi-view visualizations~\cite{roberts1998encouraging}. Such multi-view visualizations can provide a comprehensive and contextual understanding of the data in the forms of dashboards, infographics, and data stories. Due to the lack of high-quality multi-view datasets with contextual information, \rv{we carefully collected and selected 849 data stories and 249 dashboards from Calliope~\cite{shi2020calliope}, an online data story generation platform, and Tableau Public~\cite{tableau2006}, an online platform for creating and sharing data visualizations, respectively, comprising a total of 6014 visualizations. The dataset is publicly available at \url{https://chart2vec.idvxlab.com/}.} The collected dataset covers ten common topics, including economy, sports, society, health, politics, industry, \rv{recreation}, food, education, and ecology.
Third, we need to set up multiple deep learning tasks to learn the contextual information from a set of input embeddings. \rrv{We integrated both supervised and unsupervised learning tasks, where we use the linearly interpolated loss function for sequentially connected charts to learn logical associations, and introduce the triplet loss to capture the co-occurrence of the charts. Meanwhile, we employ a multi-task training strategy and optimize the results by setting hyperparameters and automatic updates.}

In this paper, we propose Chart2Vec, a model that learns a universal embedding of \rrv{visualizations}, extracts context-aware information, and enables other downstream applications such as recommendations, storytelling, and generation. To investigate the effectiveness of the proposed model, we conducted extensive evaluations including ablation studies, a user study, and quantitative comparisons with existing visualization embedding methods~\cite{zhao2020chartseer,sun2022erato}. 

In summary, the major contributions of this paper are as follows:
\begin{itemize}
    \item We collect a high-quality context-aware visualization dataset and formulate an input embedding that incorporates both factual and semantic information.
    \item We present Chart2Vec, a representation model to learn a universal embedding of \rrv{visualizations}, extract context-aware information, and enable various downstream applications.
    \item We summarize the key lessons learned during the design and development of Chart2Vec, which we hope to benefit subsequent visualization applications and related research.
\end{itemize}

%% file: sections/2relatedwork.tex
\section{Related Work}
\rv{In this section, we present a comprehensive review of the related literature, specifically focusing on representation learning in visualization, automatic multi-view visualization, and visualization similarity computation. Different from existing approaches, Chart2Vec combines structural and semantic information of \rrv{visualizations}. In addition, Chart2Vec introduces contextual relationships in multi-view visualization, a feature that can further enhance the efficiency of various downstream tasks such as recommendation, clustering, and generation.}

\subsection{Representation Learning in Visualization}
Representation learning is a machine learning technique that automatically learns representations of data~\cite{bengio2013representation}. It has been widely used in various fields, including graph learning~\cite{hamilton2017inductive}, computer vision~\cite{he2022masked}, and natural language processing~\cite{pennington2014glove}. Recently, representation learning has also been applied to address a variety of visualization tasks, such as transformation, comparison, and recommendation~\cite{wu2021ai4vis}. Since representation learning is a data-driven technique, we can divide representation learning in visualization into three categories according to the different forms of visualization data: representation learning of visualization graphics, representation learning of visualization programs, and representation learning of hybrid visualization data.

The approach of learning representations about visualization graphics focuses on extracting visual features in visualization. For example, VisCode~\cite{zhang2020viscode} extracted the visual importance map from a static visualization image and then embedded the information into the image for visualization steganography. Recent studies have mainly focused on identifying visual attributes in the visualization for subsequent visualization captioning. Lai et al.~\cite{lai2020automatic} employed the Mask R-CNN model to extract features from visual elements and visual attributes, while Qian et al.~\cite{qian2021generating} extracted information from the bounding box metadata and fused the information of the original image extracted by CNN.

Representation learning methods based on visualization programs focus on the input data as structured text and extract implicit features from the structure or text content. ChartSeer~\cite{zhao2020chartseer} utilized a syntax analyzer to convert charts in Vega-Lite specifications into one-hot vectors, which were then input into a CNN structure to obtain the representation of charts. Erato~\cite{sun2022erato} took the semantic information in the visualization as string sentences, then adopted BERT~\cite{devlin2018bert} to obtain the initial sentence vector and applied it to the visualization interpolation task by fine-tuning. 
\rrv{In addition, Draco abstracts the design rules into a set of constraints and utilizes the weights assigned to these soft constraints as feature vectors for the visualizations. To determine these weights, Draco employs RankSVM~\cite{herbrich1999ranksvm} to automatically learn their values. Consequently, the resulting vectors can be used to assess the quality of individual visualizations.}
\rrv{However, existing representation learning methods from visualization programs often serve specific tasks such as chart recommendation, generation, and evaluation}. There is still a lack of representation learning methods that can be applied to a variety of different visualization tasks based on the chart characteristics.

There are also studies for visualization representations of hybrid visualization data. For example, KG4VIS~\cite{li2021kg4vis} transformed visualizations into knowledge graphs to learn their representations. Li et al.~\cite{li2022structure} extended the tree-like format of SVGs to directed graphs and used graph convolutional neural networks to learn visualization representations.

Inspired by previous work, we propose Chart2Vec, a representation model to learn a universal embedding of \rrv{visualizations}, from declarative specifications. \rv{Chart2Vec distinguishes itself from previous approaches by incorporating both structural information introduced in ChartSeer~\cite{zhao2020chartseer} and semantic information as proposed in Erato~\cite{sun2022erato}. Furthermore, it follows the concept of Word2Vec~\cite{mikolov2013efficient} to learn the implicit features of visualization through contextual information, thereby adeptly capturing the contextual relationships among multi-view visualizations based on comprehensive multi-level chart information.}

\subsection{Automatic Multi-view Visualization}
Multi-view visualizations allow users to view a large number of data attributes and values in multiple views of visualizations coherently~\cite{chen2020composition}. Due to their capability to promote a more comprehensive understanding of data than single charts~\cite{roberts1998encouraging}, multi-view visualizations are widely used in visual analytics and visual narratives. The presentations of multi-view visualizations are mainly in the forms of dashboards~\cite{sarikaya2018we}, infographics~\cite{segel2010narrative}, and data stories~\cite{shi2020calliope}.

The rise of intelligent techniques has heightened the demand for effective visual presentation and analysis of big data systems. This has led to the emergence of automatic multi-view visualization methods, categorized into rule-based and machine learning-based approaches. 

The rule-based approaches for multi-view visualizations rely on the design guidelines from domain knowledge to automate the process. For example, DataShot~\cite{wang2020datashot} selected generated single charts through a density-based top-n algorithm, and organized multiple related visualizations into fact sheets with the same topic. Calliope~\cite{shi2020calliope} generated data stories using a logic-oriented Monte Carlo tree search algorithm with search rules derived from expert experience, while the multiple charts in each data story were arranged in a logical order. ChartStory~\cite{zhao2023ChartStory} organized multiple charts into data comics with narrative capabilities, based on established design principles. In addition, Medley~\cite{pan2023medley} recommended multiple visual collections by inferences based on a specified mapping relationship between user intent and visual elements.

Machine learning-based approaches accomplish corresponding tasks based on the results of trained models. MultiVision~\cite{wu2022multivision} calculated chart scores from chart features through an LSTM network and modeled the multiple chart selection process as a sequence-to-one regression problem. Due to the lack of datasets of multi-view visualizations, Dashbot~\cite{deng2023dashbot} utilized the reinforcement learning paradigm to simulate human behavior in exploratory visual analysis to realize the selection of charts. Erato~\cite{sun2022erato} took a new perspective by treating visualizations as semantic sentences, representing visualizations as vectors, and using them for subsequent interpolation tasks. 

The automatic multi-view visualization work mentioned above is mainly related to three major tasks: visualization recommendation, visualization clustering, and visualization generation. Chart2Vec encodes a visualization as a vector and takes into account the contextual relationship between visualizations, which can greatly improve the efficiency of the subsequent visualization tasks.

\subsection{Visualization Similarity Computation}
Visual similarity computation is a crucial process for many downstream applications, such as visualization recommendation and generation.
Currently, two major types of visualization features are considered when computing visualization similarity: textual features and graphical features~\cite{wu2021ai4vis}. Text features refer to the textual content within visualizations such as titles and captions. For example, ScatterNet~\cite{ma2018scatternet} used deep neural networks to extract semantic features from scatterplot images for similarity calculation. Vizcommender~\cite{oppermann2020vizcommender}, a content-based visualization recommendation system, leveraged the extracted text features to predict semantic similarity between two visualizations. \rv{GraphScape~\cite{kim2017graphscape} describes graphs based on the declarative grammar of Vega-lite, and calculates the similarity between graphs by the transformation cost between specifications.}

To extract graphical features for similarity computation, Demiralp et al.~\cite{Demiralp2014TVCG} estimated differences between visualizations using a perceptual kernel based on visual encoding. Li et al.~\cite{li2022structure} converted SVGs into bitmaps for visual information extraction and graphs for structural information extraction. They then applied contrastive representation learning techniques to generate embedding vectors for visual and structural information separately, which are used to calculate similarity. Additionally, Luo et al.~\cite{luo2020steerable} proposed a vector with five features to consider both the similarity score and the diversity when selecting the top-k visualizations. 
Some systems combined textual features and graphical features to improve chart detection~\cite{choudhury2016scalable} as well as classification tasks~\cite{kim2018multimodal}. For example, Kiem et al.~\cite{kim2018multimodal} proposed a multimodal model that used both pixel data and text data in a single neural network to classify the visualization. Chart Constellations~\cite{xu2018chart} measured chart similarity using visual encoding and keywords extracted from charts. To comprehensively measure the similarity between visualizations, our work considers both semantic information in text content from data attributes and structural information concerning visualization designs. 

In addition to structural and semantic information, multi-view visualizations require visual similarity metrics that consider contextual information. However, existing work has primarily focused on individual chart characteristics. Meanwhile, context-aware analysis has been applied to other data types in the visualization domain. For example, graph data are often associated with context analysis since it helps interpret and explore network structures~\cite{Mezni2022TVCG}. In the case of multi-view visualizations for tabular data, context information refers to the co-occurrence or the logical relationships~\cite{shi2020calliope} among multiple views. In this paper, we incorporate such context-aware information in our dataset collection and model design.

%% file: sections/3dataset.tex
\section{Dataset}
\rv{Given the absence of readily available high-quality context-aware visualization datasets, we collected the dataset in order to complete the training. This section provides an overview of our data collection and screening process, followed by a detailed description of the final dataset. We delve into various aspects of the dataset, including the data sources, the data filtering conditions, the final amount of data collected and the classification methods.}

\subsection{Data Collection and Screening}

Multi-view visualizations are used in many domains, including business intelligence dashboards and data stories. To prepare a high-quality dataset for model training, we searched established visualization websites and business intelligence platforms, such as Tableau Public~\cite{tableau2006}, Plotly Community Feed~\cite{plotlycommfeed}, PowerBI~\cite{powerbi}. In addition, some previous work has collected multi-view visualizations~\cite{bach2022dashboard}. \rv{Among the popular websites and platforms mentioned, we collect dashboards from Plotly and Tableau Public. Meanwhile, data story generation platforms such as Calliope~\cite{calliopedata} contain a large amount of context-aware multi-view visualizations in a factsheet format. We were able to access the chart backend data through exclusive assess to the Calliope's database.}

\rv{The visualizations on those platforms are created and edited by various users, so we still need to carefully screen all the collections. To ensure a diverse multi-view visualization dataset, we examined the coverage of various data domains in the screening process.}
Two of the authors screened and examined existing data stories on Calliope~\cite{calliopedata} for high-quality data stories. Both authors are experienced in data visualization and have conducted several data story workshops. To ensure completeness, effectiveness, and content richness, we applied the following selection rules from multiple perspectives: \rrv{(1) each multiple-view visualization must have an appropriate amount of information~\cite{gershon2001what}, with a minimum chart number set to three. All the charts in the corpus should be complete with no missing captions, data stories should not have empty titles and no duplicate charts appear in the same data story; (2) we ensure that in the same multi-view visualization, transitions between charts are related to data storytelling, where any of the six transition categories (i.e., dialogue, temporal, causal, granularity, comparison, and spatial transitions), can be discovered to maintain narrative flow and coherence~\cite{hullman2013deeper}, and (3) the charts in the same story need to be logically coherent (i.e., the content of the charts is logically connected as defined by the logicality in~\cite{shi2020calliope}).} The two authors checked all the data stories separately, marking them separately for compliance with the above criteria. If both authors approved a data story, it was added to our dataset. If they disagreed, they discussed until reaching a consensus. In the end, we selected 849 data stories.


\rv{We retrieved 10010 dashboards from Plotly Chart Studio~\cite{plotlycommfeed} using the Plotly API and also utilized the Tableau Public API to get 3176 dashboards. 
We first filtered out the dashboards containing fewer than three charts and excluded 3D visual charts. Then, we excluded those missing important data fields, which are indispensable for meaningful data \rrv{visualizations}. After the first round of screening, we collected a dataset comprising 551 qualified dashboards from Plotly and 2315 qualified dashboards from Tableau Public. Subsequently, our two authors conducted a second-round screening to assess the contextual relations between multiple charts in the same dashboard\rrv{, following the same criteria used to collect the data stories.}.
Despite recent academic research indicating an increase in narrative dashboards~\cite{sarikaya2018we}, most existing dashboards on Plotly are still mainly used for exploratory analysis, and the overall quality is not good enough to learn context-aware information. Therefore, we decided to exclude the dashboards from Plotly, and keep only data stories collected from the Calliope platform and dashboards from Tableau Public.}

\subsection{Dataset Descriptions} \label{sec:dataset-description}
After a third-round screening by another author, we obtained 849 high-quality data stories \rv{and 249 dashboards that contain 6014 visualizations.} We give detailed descriptions of the dataset classification and format.

\begin{figure}[!t]
    \centering
    \includegraphics[width=\linewidth]{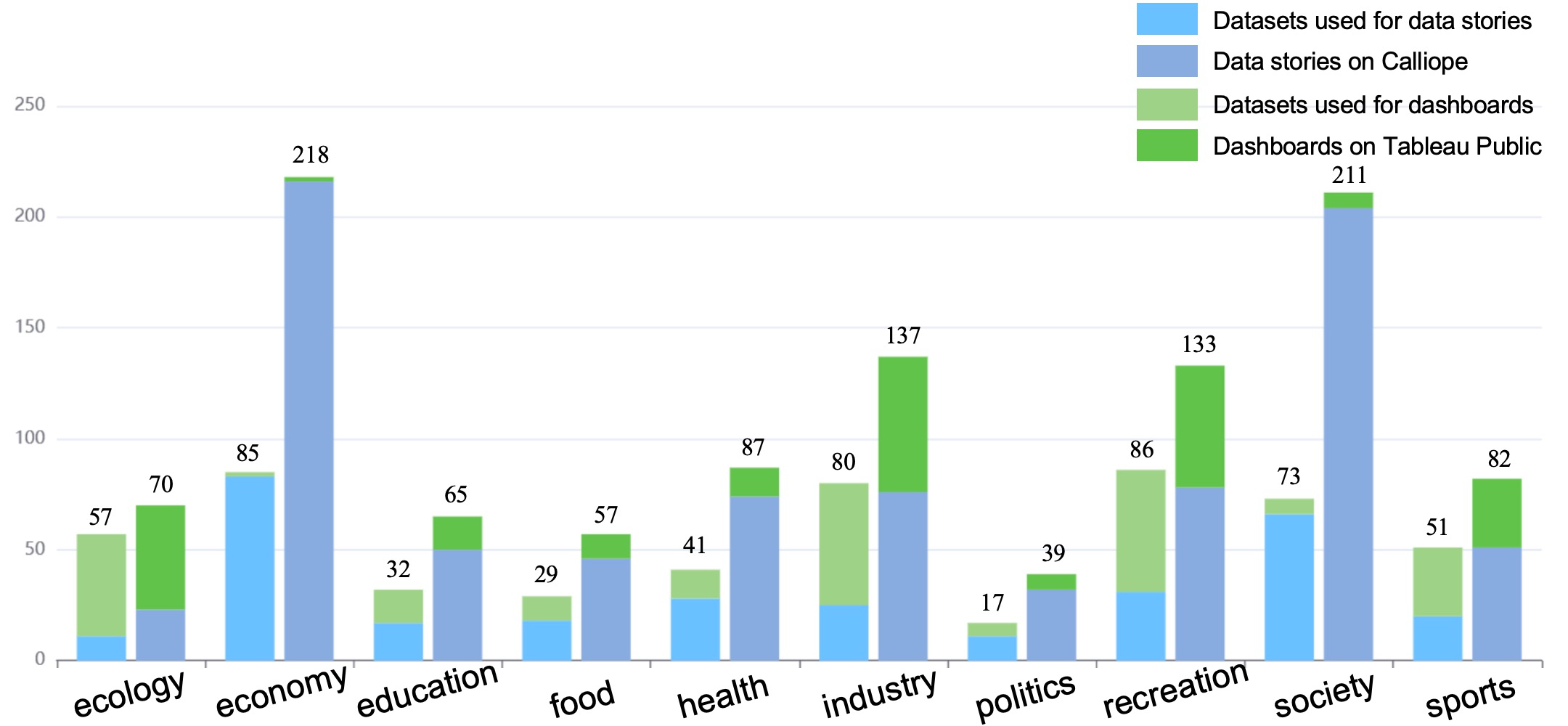}
    \caption{Distribution of datasets, data stories and dashboards in different domains.}
    \label{fig:dataset-domains} 
\end{figure}

\textbf{Dataset statistics.} Each data story contains 5-8 charts. 
The majority of data stories consisted of 5 charts, accounting for 68.9\% of the total. The 849 data stories were created from 310 datasets \rv{and the 249 dashboards are from 241 datasets} covering 10 different domains: economy, sports, society, health, politics, industry, recreation, food, education, and ecology. The classification of datasets, data stories \rv{and dashboards} are shown in Fig.~\ref{fig:dataset-domains}.

\textbf{Dataset format.} Calliope uses declarative specifications~\cite{satyanarayan2017vega} to encode each visualization in a data story, \rv{and the dashboards on Tableau Public are in the form of images, which we manually transformed into declarative specifications similar to those of Calliope for subsequent uniform processing.} We then stored individual data visualizations in the form of a list. Each item in the list corresponds to an individual chart \rv{in a multi-view visualization} and is arranged in order. The data for each chart includes the chart type, the data facts, and the column information of the raw data table.
In addition, we performed pre-processing operations on the collected data such as data completion and data correction. For example, if any column names of the \rrv{collected} dataset are abbreviated, we retrieved the URL of the \rrv{collected} dataset to identify the full names of these abbreviations and then replaced them manually. We also corrected the misspelled column names.

%% file: sections/4model.tex
\section{Methodology}
This section introduces the design and implementation of Chart2Vec. The goal is to convert the visualization into an embedding vector through the Chart2Vec model. The vector representation not only retains meaningful information (i.e., structural and semantic information) about individual charts but also captures the contextual relationships between charts. We first define the form of a chart and then describe the implementation details of the Chart2Vec model. \rv{We also open-sourced the training and testing data along with our trained model on GitHub at} \url{https://github.com/idvxlab/chart2vec}.

\subsection{Chart Characterization}
\rv{The goal is to learn a universal embedding of \rrv{visualizations} from declarative specifications. To achieve this, it \rrv{is crucial} to establish a declarative syntax format for the charts. This format is required to effectively represent the essential information of the charts.} Inspired by recent work in natural language understanding and automatic generation of semantic content from visualizations~\cite{lundgard2021accessible,srinivasan2018augmenting}, we aim to characterize a format that is more general and comprehensive than existing representations. According to the model proposed by Lundgard \& Satyanarayan~\cite{lundgard2021accessible}, semantic content from visualizations through natural language descriptions can be classified into four levels: elemental and encoded properties (L1), statistical concepts and relations (L2), perceptual and cognitive phenomena (L3) and contextual and domain-specific insights (L4). In the following, we describe the process of constructing the \rv{declarative specification of a chart} in conjunction with the four-level semantic model.

As mentioned in Section~\ref{sec:dataset-description}, we utilized the curated \rv{multi-view visualizations as our training and testing datasets and manually transformed the dashboards on Tableau Public into declarative specifications consistent with that in Calliope}. Each \rv{ multi-view visualization} contains a set of interconnected individual charts that convey meaningful insights. Calliope employs ``data facts'' to store the essential information of the chart, which concentrates on capturing the data content from a semantic perspective. Each data fact can be expressed as a 5-tuple: \rrv{$f_i=\left \{ \text{\textit{type, subspace, breakdown, measure, focus}}
\right \}$}. However, the data fact definition in Calliope solely includes the L2 information related to the aforementioned semantic model.

To enhance the richness of chart content, we improve the original form by incorporating the chart type and meta information. The chart type, such as bar chart or scatterplot, is indispensable from the visual encoding perspective and corresponds to the L1 information. Meta information describes the perceptual and cognitive features of the visualization and thus corresponds to the L3 information. For instance, if a chart represents a trend, we additionally describe whether the trend is ascending or descending.

After making the aforementioned refinements, we introduce a more comprehensive chart representation, which we refer to as a ``chart fact''. It is defined as a 7-tuple:
\begin{equation}
\begin{aligned}
c_i &= \left \{ \text{\textit{type\textsubscript{c}, type\textsubscript{f}, subspace, breakdown, measure, }} \right.
 \text{\textit{focus, meta}} \} \\
&= \left \{ ct_i, ft_i, s_i, b_i, m_i, f_i, e_i\right \} \nonumber
\end{aligned}
\end{equation}
where \textbf{\textit{type\textsubscript{c}}}
(denoted as $ct_i$) indicates the type of chart and
\textbf{\textit{type\textsubscript{f}}}
(denoted as $ft_i$) expresses the type of information described by the data fact. Similar to Calliope, we support 15 commonly used chart types and 10 data fact types; \textbf{\textit{subspace}} (denoted as $s_i$) comprises a set of data filters that can select a specific range of data, defined as $\left \{ \left \{\mathcal{ F}_1=\mathcal{V}_1  \right \} , ..., \left \{ \mathcal{ F}_k=\mathcal{V}_k \right \}  \right \}$, where $\mathcal{F}_i$ denotes the data field and $\mathcal{V}_i$ denotes a corresponding specific value in $\mathcal{F}_i$; \textbf{\textit{breakdown}} (denoted as $b_i$) consists of a single temporal or categorical data field that can further divide the data items into groups in the subspace; \textbf{\textit{measure}} (denoted as $m_i$) is a numerical data field that can be used to measure the data in the groups through different aggregation methods; \textbf{\textit{focus}} (denoted as $f_i$) indicates the data item or data group that requires attention; \textbf{\textit{meta}}  (denoted as $e_i$) indicates additional information about the chart. The meta field contains different information depending on the fact type, as described in detail in Table~\ref{tab:model-meta}. 

\begin{figure*}[!t]
    \centering
    \includegraphics[width=2\columnwidth]{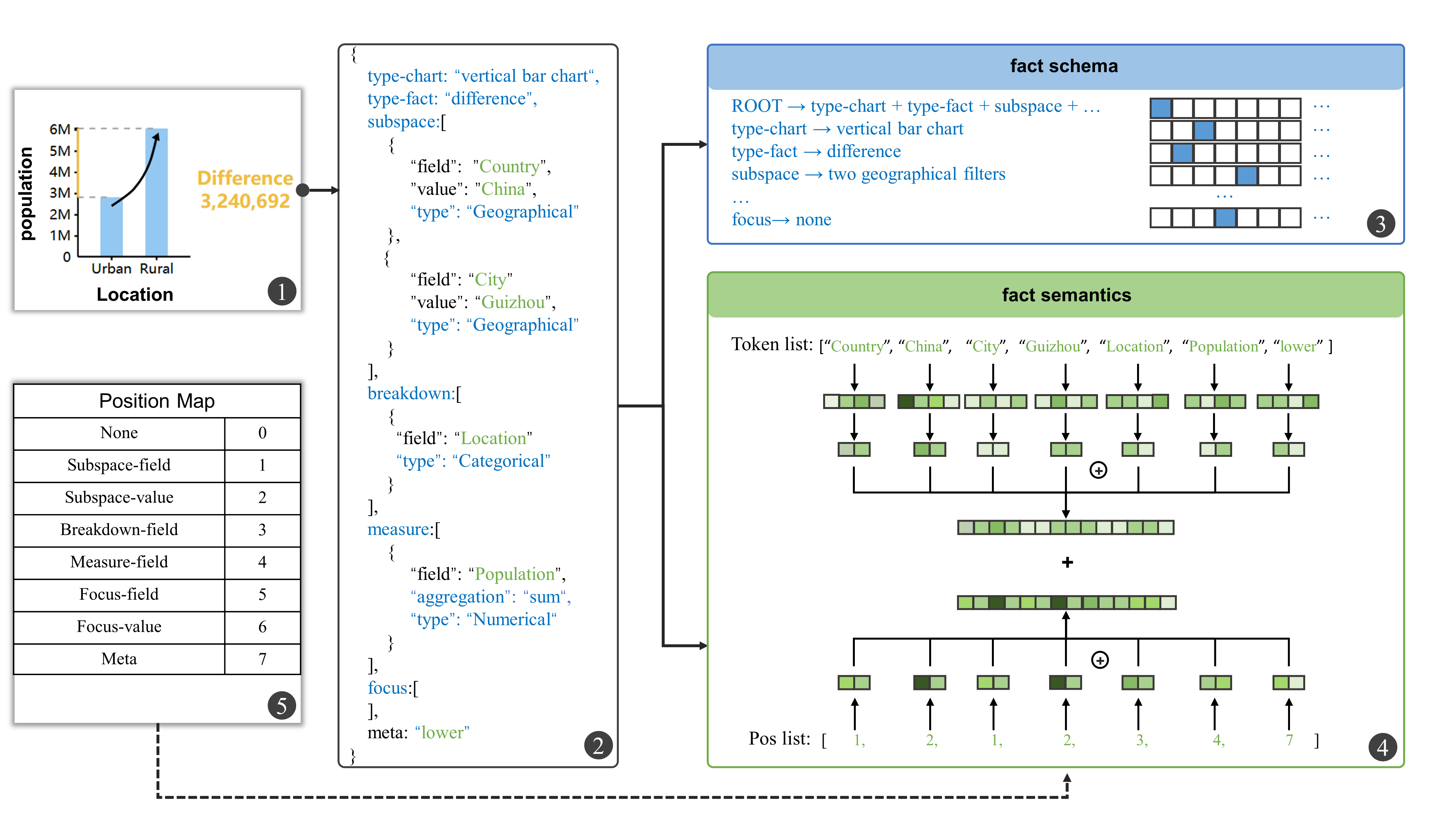}
    \caption{The formulation details of an example chart fact: (1) the graphical presentation of the visualization data, (2) the example chart fact representation,  (3) the fact schema which shows structural information in the chart fact, (4) the fact semantics which indicates semantics information in the chart fact, and (5) the location of the fields in the chart fact where stores semantic information.}
    \label{fig:model-chart-info} 
\end{figure*}

To provide a better understanding of the 7-tuple and its correspondence with the chart content, we present a concrete example. Consider a dataset containing useful information about schools in the world, including columns such as school name, location, and the number of students. Suppose we generate a chart from the dataset that depicts the difference in student numbers between rural and urban areas in Guizhou, China, represented as a vertical bar chart as shown in Fig.~\ref{fig:model-chart-info}(1). The corresponding chart fact is shown in Fig.~\ref{fig:model-chart-info}(2) as \{``vertical bar chart'', ``difference'', \{\{Country =``China''\}, \{City=``Guizhou''\}\}, \{Location\}, \{sum(Population)\}, ``lower''\}.

\begin{table}[!ht]
\small
\caption{The meta information for different fact types in the 7-tuple.}
 \centering
 \resizebox{\columnwidth}{!}{
 \begin{tabular}{ccccc}
  \toprule
  \makebox[0.1\columnwidth][c]{\textbf{fact type}} & \makebox[0.1\columnwidth][c]{\textbf{meta information}} \\
  \midrule
  trend  & \makecell[l]{The overall direction of the trend.\\ The options are ``increasing'', ``decreasing'' and ``no trend''.}\\
  \midrule
  categorization  &  \makecell[l]{The total number of categories. For example, if the category \\has 20 categories, then the meta value is ``20 categories''.} \\
  \midrule
  difference  &  \makecell[l]{The difference between the two values. For example, as shown in\\ Fig.~\ref{fig:model-chart-info}(1), if the urban value is lower than the rural value, the\\ meta is ``lower''; otherwise, it is ``higher''.}\\
  \midrule
  rank  &  \makecell[l]{Top three ranking values. For example, if a chart is a ranking\\ of total car sales, the meta value is the top three car brands.}\\
  \midrule
  extreme  &  \makecell[l]{The types of the extreme. \\The options are ``max'' and ``min''.}\\
  \midrule
  association  &  \makecell[l]{The type of association. \\The options are ``positive'' and ``negative''.}\\
  \bottomrule
 \end{tabular}
 }
 \label{tab:model-meta}
\end{table}

A chart fact contains both structural and semantic information, with structural information defined as the \textbf{\textit{fact schema}} and semantic information as the \textbf{\textit{fact semantics}}. Chart fact consists of a 7-tuple and the \textbf{\textit{fact schema}} can be categorized following the detailed principles: (1) the options for \textit{$type_{c}$} and \textit{$type_{f}$} are fixed and enumerable; (2) the number of filters in the \textit{subspace} can be none, one, or more than one; (3) the value of \textit{breakdown} can either perform a grouping action or not; (4) the aggregation methods of \textit{measure} include count, sum, average, minimum, and maximum; (5) the \textit{focus} and the \textit{meta} can either have additional information or not; (6) additionally, the field types in subspace, breakdown, measure, and focus are one of four fixed types: temporal, numerical, categorical, and geographical. The structure follows the context-free grammar (CFG)~\cite{hopcroft2006introduction}, which is a set of recursive rules used to generate patterns of strings. Therefore, the structural information of each chart can then be represented as a parse tree generated by a set of rules within the CFG, and we give all the rules of the chart fact in the supplementary material. This transformation facilitates the subsequent encoding of structural information, as explained in Section~\ref{sec:input-emb}. The \textbf{\textit{fact semantics}} refers to the semantic content within the chart fact, including the information from the data field and the value. Since the fact semantics are highly related to the dataset, they are not enumerable and their data semantics are mostly inconsistent. In Fig.~\ref{fig:model-chart-info}(2), the text highlighted in blue represents structural information, while the text highlighted in green represents semantic information. The above structural information can be represented as the rules shown on the left side of Fig.~\ref{fig:model-chart-info}(3). The example fact semantics include ``Country'', ``China'', ``City'', ``Guizhou'', ``Location'', ``Population'', ``lower'', etc. The semantic information is then organized into a token list, as shown in Fig.~\ref{fig:model-chart-info}(4).


\subsection{Chart2Vec Model}
To construct a universal embedding of charts, we not only incorporate chart information at different levels but also define multiple tasks to learn the vector representation of the chart. This section begins with an overview of the model inputs and architecture, followed by the implementation details and model training configurations.

\subsubsection{Overview}
In order to better understand the inputs and outputs of the model, we provide a formal definition and an overview of the overall architecture. 

\begin{figure}[!t]
    \centering
    \includegraphics[width=0.9\linewidth]{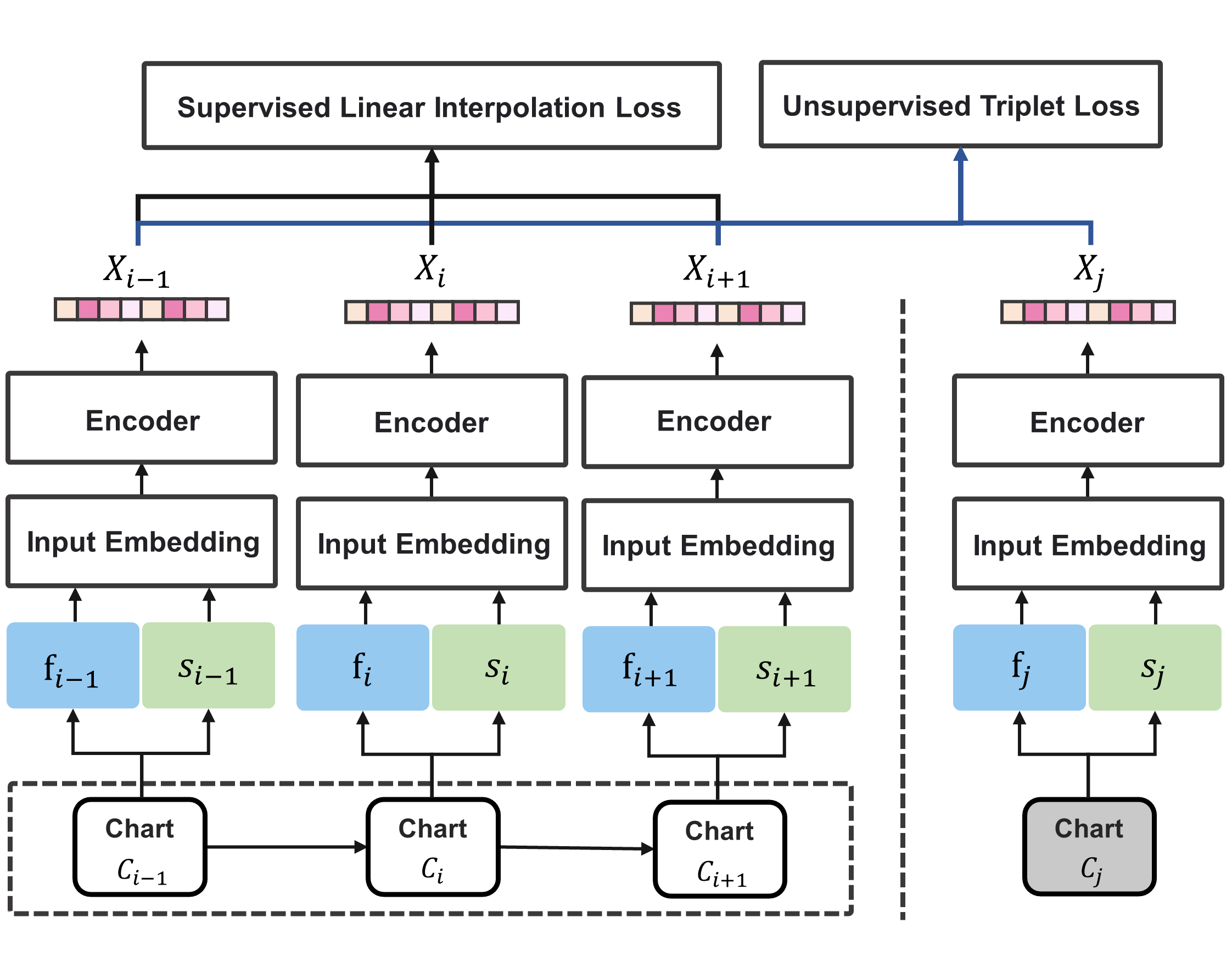}
    \caption{Formulation of Chart2Vec. During model training, the inputs are a set of four visualizations, which are passed through the Chart2Vec model with shared parameters. The two loss functions are used to jointly optimize the model parameters.}
    \label{fig:model-input-format} 
\end{figure}

\textbf{Formulation.} \label{sec:formulation}
In the proposed model, a single chart denoted as $C_i$ is taken as input, and a corresponding output vector denoted as $X_i$ is generated. Each input vector consists of two types of information: fact schema and fact semantics, represented by $f_i$ and $s_i$ respectively. To ensure that contextual correlations are captured by the vector representation, we use an input set of four charts that are fed into the same model with shared parameters. Each input set consists of three sequentially connected charts in the same multi-view visualization, denoted as $\left \{ C_{i-1}, C_i, C_{i+1}\right \}$, as well as a random chart from other multi-view visualizations generated, denoted as $C_j$. Therefore, each input set can be represented as $\left \{ C_{i-1}, C_i, C_{i+1}, C_j\right \}$, and the corresponding output vector is $\left \{ X_{i-1}, X_i, X_{i+1}, X_j\right \}$. More details can be found in Fig.~\ref{fig:model-input-format}.

\begin{figure}[!t]
    \centering
    \includegraphics[width=0.7\linewidth ]{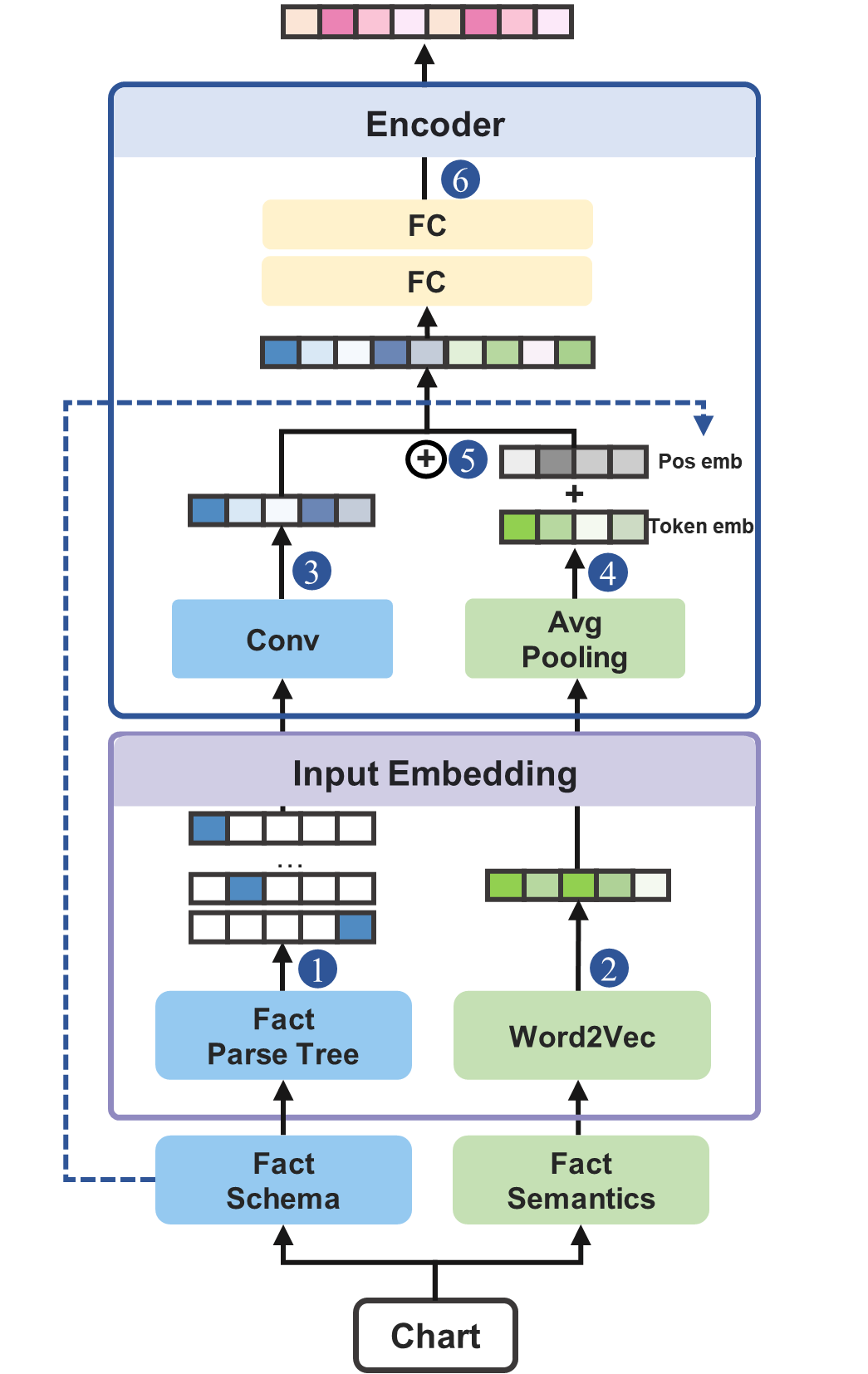}
    \caption{Architecture of the Chart2Vec model. 
}
    \label{fig:model-architeture} 
    \vspace{-1em}
\end{figure}

\textbf{Architecture.} The Chart2Vec model architecture, as illustrated in Fig.~\ref{fig:model-architeture}, comprises two main components: an input embedding module and an encoder. \rv{The input embedding module is designed to convert the chart fact format into a \rrv{numeric} form that can be \rrv{calculated} by the computer, and it needs to be able to extract as much valid information as possible from the original format of the chart. The original format of the chart can be seen as composed of two parts, namely fact schema and fact semantics.} The fact schema is represented in a rule tree format and converted into one-hot vectors \rrv{as it consists of enumerable properties} (Fig.~\ref{fig:model-architeture}(1)), while the fact semantics is encoded using the Word2Vec model~\cite{mikolov2013efficient}, with each word encoded individually \rv{to represent its semantic information}(Fig.~\ref{fig:model-architeture}(2)). \rvv{To perform information fusion and vector space conversion,} the encoder module convolves the one-hot matrix corresponding to the fact schema (Fig.~\ref{fig:model-architeture}(3)) and averages the word vectors encoded by the fact semantics (Fig.~\ref{fig:model-architeture}(4)), followed by fusion operations on the averaged word vectors, the position encoding corresponding to the fact schema structure, and the one-hot matrix after convolution (Fig.~\ref{fig:model-architeture}(5)). \rrv{Finally, to enhance the model's accuracy and generalization capability, the resulting vector is obtained by passing the concatenated vector through two fully connected layers for nonlinear transformation. Moreover, the necessity of adding fully connected layers is validated in Section 5.1.}
\subsubsection{Input Embedding} \label{sec:input-emb}
\rrv{The chart fact exists in the form of structured strings. With the input embedding module, it is possible to transform the raw data into a form that can be understood and processed by the model. In addition, since the fact consists of two parts, fact schema and fact semantics, each with distinct different attributes and features, we encode them separately with different methods to effectively represent the information so that the model can understand and process them more accurately.}

\textbf{Input embedding of fact schema.} \rrv{As described in Section 4.1, the structural information in the fact schema can be represented as a parse tree generated by a set of rules in CFG. Therefore,} we developed a rule tree in a CFG format, containing \rv{60 rules}, with each rule encoded using a \rv{60-dimensional} one-hot vector. As the maximum number of rules corresponding to the structural information of a chart is 16, the input embedding of the structural information in each fact can be represented as a vector matrix of size 16$\times$\rv{60}, as depicted in Fig.~\ref{fig:model-architeture}(1).

\textbf{Input embedding of fact semantics.} Fact semantics are field values in the chart fact that contain semantic information, which are derived from column names or values in the original dataset. As each field value has its specific meaning, it cannot be represented by fixed rules. Fig.~\ref{fig:model-chart-info}(5) shows that there are seven fields in the chart fact that contain semantic information: subspace-field, subspace-value, breakdown-field, measure-field, focus-field, focus-value, and meta.  First, we extract the relevant words from the seven fields, excluding any field with an empty value. It is worth noting that each field may contain multiple words, which are subsequently split into separate words. For example, consider the original list of extracted words, which includes [``Country name'', ``City name'', ``Year'', ``Student population'', ``Year'', ``2018'']. After finer granularity segmentation, we obtain a list of individual words, including [``Country'', ``name'', ``City'', ``name'', ``Year'', ``Student'', ``population'', ``Year'', ``2018'']. \rvv{However, since the model cannot \rrv{calculate} string data directly, we need to convert the words into numerical inputs. To preserve the meaning of each word, we use the Word2Vec model to transform each word into a vector}, as illustrated in Fig.~\ref{fig:model-architeture}(2). 
In this paper, we employ the \rv{pre-trained model provided by Wikipedia2Vec~\cite{yamada2020wikipedia2vec,yamada2016joint}} \rvv{to benefit from reduced training time, enhanced encoding of word vectors, and improved overall performance.}

\subsubsection{Encoder}
After obtaining the initial vectors for both fact schema and fact semantics through the input embedding module, the encoder module employs two distinct operations, namely feature pooling and feature fusion, to achieve the final vector representation. \rrv{Among them, feature pooling is designed to extract the most important features and reduce the computational complexity of the model by minimizing the feature dimensions. Feature fusion can be used to combine feature information from different parts, facilitating the interaction and information transfer between different features to improve the richness and expressiveness of features.}

\textbf{Feature Pooling.}
We utilize convolutional layers to extract the fact schema features and transform the one-hot vectors into a single vector for the fact schema part (Fig.~\ref{fig:model-architeture}(3)). \rvv{This choice is motivated by the ability of convolutional layers to capture local relationships among rules in their order. Additionally, prior research~\cite{zhao2020chartseer} has demonstrated that CNNs outperform RNNs in encoding visualization data embeddings. This could be attributed to the repetitive nature of the input CFG rules that rely on declarative programs, which are perceived as translationally invariant substrings.} 
For the fact semantics part, we apply an averaging pooling operation to each word vector by averaging the original word vector in fixed intervals (Fig.~\ref{fig:model-architeture}(4)). For example, suppose a word is initially represented by a 100-dimensional vector. We can transform it into a 10-dimensional vector by applying the average pooling operation with a step size of 10. \rvv{This operation enables us to capture the topic information of each word while blurring its boundaries. Additionally, it reduces the size and computational cost of feature fusion. To confirm the effectiveness of this strategy, we perform ablation studies in the evaluation section to assess whether it indeed enhances the model's performance.}


\textbf{Feature Fusion.}
We extracted both structural and semantic information from the 7-tuple chart fact. \rvv{While the semantic part is derived from the extracted words in the field values of the chart fact, the connection between the structural and semantic information is lost during the extraction process. To restore the lost connection}, we introduce location markers to indicate where the semantic information is extracted from the original chart fact (as shown in Fig.~\ref{fig:model-chart-info}(5)). We add each word with its corresponding location number and concatenate the structural vectors and semantic vectors with the added location information (as shown in Fig.~\ref{fig:model-chart-info}(4)). Finally, a double-layer fully connected layer is employed to perform a nonlinear transformation, resulting in the final vector representation (as shown in Fig.~\ref{fig:model-architeture}(6)).

\subsubsection{Loss Function}
To better learn the contextual associations between charts, we adopt a multi-task training strategy that combines supervised and unsupervised learning tasks to optimize the loss function.

\textbf{Supervised Linear Interpolation Loss.} Inspired by Erato~\cite{sun2022erato}, we employ a linear interpolation loss function \rvv{for the three consecutive charts to establish the final vector that captures contextual relationships}:
\begin{equation}
    \begin{aligned}
       l_1 =\sum_{k=1}^{D} \left ( d\left ( C_{k_i} ,C_{k_{mid}}  \right ) + \alpha \sum_{t,s\in \left \{ i-1,i,i+1 \right \} }^{t\ne s} d\left ( C_{k_t} ,C_{k_s} \right )   \right ) 
    \end{aligned}
    \label{eqn:loss1}
\end{equation}
where $k$ denotes a training sample, and $D$ is the total number of training sets. $d\left ( \cdot  \right )$ represents the Euclidean distance between two vectors. The equation consists of two parts. $C_{k_i}$ in the first part is the $C_i$ mentioned in Section~\ref{sec:formulation}, which is the chart in the middle of the three sequentially connected charts in a training set. $C_{k_{mid}}$ represents the middle point obtained by linear interpolation of $C_{k_{i-1}}$ and $C_{k_{i+1}}$, i.e., $C_{k_{mid}}=\left ( C_{k_{i-1}}+C_{k_{i+1}} \right ) /2$. By minimizing the distance between $C_{k_i}$ and $C_{k_{mid}}$, the three connected charts can have a linear relationship in vector space. The second part of the formula aims to minimize the distance between the three sequential charts in the training sets. The coefficient $\alpha$ is used to balance the two parts of the equation.

\textbf{Unsupervised Triplet Loss.} \rvv{To enhance the co-occurrence relationship}, we employ an unsupervised learning task, in which input triples $C_{i-1}$, $C_{i+1}$, and $C_j$ are utilized as anchor, positive, and negative samples, respectively. By minimizing the distance between the anchor and the positive samples while maximizing the distance between the anchor and the negative, charts that appear in the same multi-view visualization can be brought closer together. The triplet loss function is defined as follows:
\begin{equation}
    \begin{aligned}
        l_2 =\sum_{k=1}^{D}\left [ d\left ( C_{k_{i-1}},C_{k_{i+1}}\right ) - d\left ( C_{k_{i-1}},C_{k_j} \right ) +m \right ]_+  
    \end{aligned}
    \label{eqn:loss2} 
\end{equation}
where $m$ is a margin factor that controls the minimum distance between the anchor and a negative sample, \rv{obtained through experimental tuning}. Specifically, the first part of the equation computes the distance between the two charts in the front and back of the three sequentially connected charts in a training sample, while the second part computes the distance between the anchor chart $C_{i-1}$ and the negative sample $C_j$. Notably, to ensure proper parameter optimization, the overall loss is constrained to be greater than or equal to zero, as the subtraction operation may result in negative values.

\textbf{Multi-task learning.} \rrv{To enable the model to learn the spatial linear logical relationships between multiple charts and cluster vectors located between the same multi-view visualizations to capture the co-occurrence relationships,} we combine the above two tasks and use a multi-task training strategy to optimize the proposed supervised linear interpolation loss (Eq.~\ref{eqn:loss1}) and unsupervised triplet loss (Eq.~\ref{eqn:loss2}):
\begin{equation}
    \begin{aligned}
       \mathcal{L}=l_1+\beta l_2
    \end{aligned}
    \label{eqn:loss3} 
\end{equation}
where $\beta$ is a hyperparameter set to balance the two loss functions. \rv{We examined the values of these two parts of the loss to see their data levels and relative sizes, and performed hyperparameter tuning based on the differences.}

\subsubsection{Model Training} \label{sec:model-training}
We describe the training corpus and configurations in detail.

\textbf{Training Corpus.} \rv{We collected 1098 multi-view visualizations from 551 datasets. 310 of these 551 datasets were sourced from Kaggle~\cite{kaggle} and the others were uploaded by users themselves, covering 10 common topics.} Among them, we randomly selected \rv{994 multi-view visualizations based on 501 datasets as the training set, and the other 104 multi-view visualizations from 50 datasets as the testing set.} Three sequentially connected charts in the same multi-view visualization are regarded as a positive triplet. Each triplet, denoted as $\left \{ C_{i-1}, C_i, C_{i+1}\right \}$, is a set of contextually related charts centered around the middle chart. For each positive triplet, negative instances are created by selecting charts from other multi-view visualizations. Each training sample consists of one positive triplet of three connected charts in the same multi-view visualization and one negative instance from a different multi-view visualization. We removed duplicate negative instances and obtained a total of \rv{42,222} training samples.

\textbf{Configuration details.} 
Based on the properties of the training set, we set the maximum length of the semantic part to 25 words. The structural part of the model employs a 3-layer CNN encoder featuring a kernel size of 3, producing a chart vector with a dimension of 540. \rv{We also use batch normalization in each layer of the model.} The model was trained using the PyTorch framework, with an initial learning rate of 0.01 and a batch size of 128. The dropout rate of the hidden layer was set to 0.1, and the parameters were updated using the Adam optimizer. \rv{The model was trained on an Nvidia Tesla-V100 (16 GB) graphics card. We conducted training for 10 epochs, comprising 3,298 steps in total, which took approximately 27 minutes to complete. Throughout the training phase, the memory consumption amounted to 1241 MiB.}


%% file: sections/5evaluation.tex
\section{Evaluation}
To assess the effectiveness of the Chart2Vec model, we performed three experiments: (1) an ablation study to demonstrate the essentiality of each module integrated into the model, (2) a user study to evaluate the coherence between Chart2Vec embedding vectors and human cognition, and (3) a quantitative comparison to gauge the performance of our approach and other chart embedding methods in capturing contextual relationships among charts.

\subsection{Ablation Study} \label{sec:ablation-study}
To thoroughly investigate the significance and indispensability of each module in the Chart2Vec model, we performed ablation experiments, concentrating on four aspects: training tasks, feature content, pooling strategy, and fusion strategy. In this section, we offer a comprehensive description of the dataset and the evaluation metrics, and present the experimental outcomes for different combinations.

\subsubsection{Dataset}
In Section~\ref{sec:model-training}, we presented the training and implementation details of the Chart2Vec model and outlined our process of collecting \rv{1098 high-quality multi-view visualizations}, out of which 104 were set aside as the test set. We extracted all the charts from the \rv{104 multi-view visualizations} in the chart fact format, resulting in a total of \rv{551} test samples of visualizations.

\subsubsection{Metrics}
After completing the aforementioned steps, we fed the \rv{551} test samples into our model to obtain their vector representations, and computed the Euclidean distance between charts within the same dataset. For each anchor chart, we selected the closest chart from the same dataset \rrv{(i.e., the chart with the smallest Euclidean distance), indicating the closest distance in the vector space. This process resulted in 551 pairs of charts.} To evaluate Chart2Vec's ability to encode contextual relationships between multiple charts, we utilized three metrics: top-2 retrieval accuracy, top-3 retrieval accuracy, and co-occurrence metric. These metrics range from 0 to 1, with higher values indicating better performance. We provide detailed explanations for each metric, as well as their calculation procedures.

\textbf{Top-2 retrieval accuracy.} For each anchor chart, we search for the nearest chart represented by its vector based on the distances between chart vectors. If two charts belong to the same multi-view visualization and are connected by less than 2 consecutive charts, the anchor chart is considered to meet this criterion. We calculate the retrieval results for all \rv{551} charts and report the percentage of charts that satisfy this criterion as the final result.

\textbf{Top-3 retrieval accuracy.} Similar to the calculation method of top-2 retrieval accuracy, we use the Euclidean distance to search for the closest chart vector to the anchor chart. For the anchor chart to meet the criteria of the top-3 retrieval accuracy metric, the retrieved chart must belong to the same multi-view visualization as the anchor chart and be connected by no more than 3 consecutive charts.

\textbf{Co-occurrence.} The co-occurrence metric measures the ability of Chart2Vec to capture the relationships between charts that frequently occur in the same multi-view visualization. To determine if an anchor chart meets the requirements for this metric, we check if the retrieved chart and the anchor chart occur in the same multi-view visualization. The final value of this metric is calculated by counting all the charts that meet the requirements and dividing by the total number of charts.

\subsubsection{Results}
We conducted experiments on 10 combinations of modules, using the same training data as Chart2Vec. These 10 combinations can be classified into four categories based on the position and function of each module in the model.

\textbf{Training tasks (2 combinations).} To demonstrate the benefits of employing multi-task joint training, we evaluated models that utilized only one of the training tasks. \textbf{\textit{No Linear Interpolation}} indicates that only the unsupervised learning task was used, and triplet loss was used to classify charts that are not in the same multi-view visualization. \textbf{\textit{No Classification}} indicates that only the linear interpolation supervised learning task was used, and the linear interpolation loss was employed to capture correlations between charts within the same multi-view visualization.

\textbf{Feature content (2 combinations).} The Chart2Vec model captures two key aspects of chart information: fact schema and fact semantics. To demonstrate the importance of each aspect, we separately removed one of them. \textbf{\textit{No fact schema}} indicates that only the semantic information of the chart was considered, while \textbf{\textit{No fact semantics}} indicates that only the structural information of the chart was considered.

\textbf{Pooling strategy (4 combinations).} To enhance the representation of chart vectors, Chart2Vec employs a pooling strategy to aggregate the initial single-word vectors obtained from the semantic component, which captures semantic thematic information. In this study, four different pooling strategies were evaluated: \textit{no word pooling}, \textit{words avg pooling}, \textit{word max pooling}, and \textit{words max pooling}. \textbf{\textit{No word pooling}} means that no pooling strategy is used and the obtained multiple-word vectors are directly concatenated and inputted into the encoder module. \textbf{\textit{Words avg pooling}} denotes that all words are averaged for obtaining the overall semantic feature. \textbf{\textit{Word max pooling}} and \textbf{\textit{Words max pooling}} represent the maximum pooling strategy applied to a single word and to all words, respectively. For example, when conducting maximum pooling on a single-word vector, a window size of 10 is applied, and the highest value among 10 values is selected as the value for the entire window. When performing maximum pooling on all words, all positions of the words are simultaneously scanned, and the maximum value at each position is selected as the value for that position.

\textbf{Fusion Strategy (2 combinations).} We consider the fusion of fact schema and fact semantics in the Chart2Vec model and validate the necessity of using a fusion strategy. Specifically, we evaluate two combinations: \textbf{\textit{No pos}} removes the positional encoding from the fact semantics and \textbf{\textit{No FC}} directly concatenates the fact schema and fact semantics with positional encoding as the final output without the addition of the fully connected layer.

The results are presented in Table~\ref{tab:ablation-study}, which demonstrate that Chart2Vec achieves superior performance with a \rv{top-2 ratio of 0.63, top-3 ratio of 0.73, and co-occurrence value of 0.81.} Removing any of the tasks leads to performance degradation. Among the different training tasks, the supervised learning task with linear interpolation has a greater impact on the model's performance. The influence of different feature contents on the model's performance varies, with fact schema being more important than fact semantics. The performance differences among the various pooling strategies are negligible, and all are inferior to the word vector pooling strategy proposed in this paper. This observation also applies to the fusion strategy.

\begin{table}[!ht]
\caption{The results of ablation study.}
\renewcommand\arraystretch{1.35}
\resizebox{\columnwidth}{!}{
\begin{tabular}{lccccc}
\hline
                        & \begin{tabular}[c]{@{}c@{}}Top-2 Retrieval\\Accuracy\end{tabular} & \begin{tabular}[c]{@{}c@{}}Top-3 Retrieval\\Accuracy\end{tabular} & Co-occurrence    & \begin{tabular}[c]{@{}c@{}} Memory\\consumption  \end{tabular}  &\begin{tabular}[c]{@{}c@{}} Time\\consumption  \end{tabular} \\ \hline
\textbf{Chart2Vec} & \textbf{0.63}  & \textbf{0.73}  & \textbf{0.81} & 1241 MiB & 27min15s                 \\ \hline
\textit{Training Tasks}          & \multicolumn{1}{l}{}                                               & \multicolumn{1}{l}{}                                     & \multicolumn{1}{l}{} \\
No Linear interpolation & 0.55   & 0.63   & 0.70    & 1241 MiB  & 24min49s           \\
No Classification       & 0.43   & 0.51   & 0.56    & 1241 MiB  & 24min17s             \\ \hline
\textit{Feature Content}        & \multicolumn{1}{l}{}                                               & \multicolumn{1}{l}{}                                               & \multicolumn{1}{l}{} \\
No fact schema          & 0.41    & 0.50    & 0.54  & 905 MiB  & 24min59s                \\
No fact semantics        & 0.53  & 0.65    & 0.74   & 1221 MiB  & 08min32s               \\ \hline
\textit{Pooling Strategy}        & \multicolumn{1}{l}{}                                               & \multicolumn{1}{l}{}                                               & \multicolumn{1}{l}{} \\
No word pooling         & 0.56   & 0.68    & 0.75   & 1523 MiB  & 25min40s               \\
Words avg pooling       & 0.59   & 0.70    & 0.77   & 1239 MiB  & 14min31s              \\
Word max pooling        & 0.59   & 0.67    & 0.74   & 1241 MiB  & 24min17s              \\
Words max pooling       & 0.56   & 0.67    & 0.74   & 1239 MiB  & 16min27s              \\ \hline
\textit{Fusion Strategy}         & \multicolumn{1}{l}{}                                               & \multicolumn{1}{l}{}                                               & \multicolumn{1}{l}{} \\
No pos                  & 0.60   & 0.72    & 0.79   & 1241 MiB  & 24min22s              \\
No FC                   & 0.49   & 0.59    & 0.64   & 1071 MiB  & 24min16s              \\ \hline
\end{tabular}
}
 \label{tab:ablation-study}
\end{table}

\subsection{User Study}
We conducted a user study to validate the effectiveness of the chart embeddings generated by Chart2Vec. The purpose of the study is to assess the consistency of the calculated similarities between the Chart2Vec embedding vectors with human perception and cognition.
\begin{figure}[!t]
    \centering
    \includegraphics[width=0.75\columnwidth]{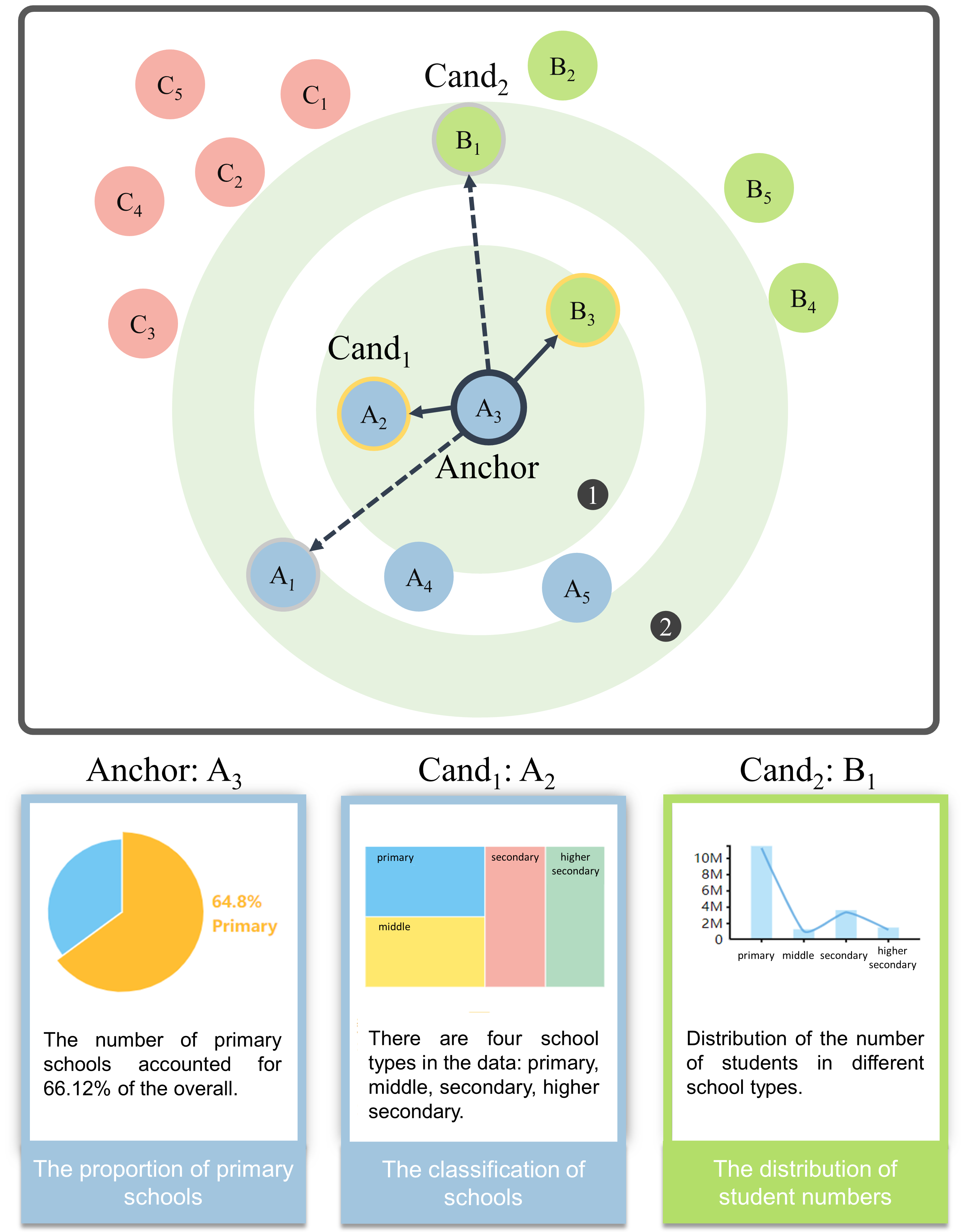}
    \caption{The construction of the user study training dataset. We selected an example dataset and represented all its related charts in this figure. Each chart is represented as a node and the charts from different multi-view visualizations are marked in different colors. Assuming $A_3$ as the anchor chart, we calculate its distance from the other charts, respectively. Two charts $A_2$ and $B_3$ fall in the range of the 15\% nearest charts shown in the filled circle \ding{182}. We randomly select one as one candidate $Cand_1$. Another two charts $A_1$ and $B_1$, locate in the second range shown in the filled circle \ding{183} and we also randomly select one as the other candidate $Cand_2$.}
    \label{fig:user-study} 
\end{figure}

\subsubsection{Dataset}
To ensure a fair and comprehensive comparison, we created 30 multi-view visualizations for this experiment, based on 10 datasets of distinct domains. We first encoded all the charts as vectors using the Chart2Vec model. Then, we randomly selected three anchor charts for each dataset, constructing a total of 30 anchor charts. For each anchor chart, we calculated its Euclidean distance with all the other charts from the same dataset. To assess whether the participants could tell the differences between the most similar charts and the moderately similar charts, we selected two candidate charts from two different similarity distance ranges. The first candidate ($Cand_1$) was selected from the top 15\% nearest charts to the anchor chart, and the second candidate ($Cand_2$) was selected from the 40\% to 50\% nearest charts. As shown in Fig.~\ref{fig:user-study}, we obtained a set of 3 charts consisting of one anchor chart ($A_3$) and two candidate charts ($A_2$ and $B_1$) as one example in the user study. This process resulted in 30 sets of charts for the experiment. To ensure that the participants could understand the meaning of the charts, we added captions that translated the chart facts into natural language descriptions for each chart \rv{and do not add any subjective narratives}.

\subsubsection{Procedure}
We recruited 30 participants (18 females, aged between 20 to 45, $mean_{age}$ = 23.8) to take part in our study. They come from diverse professional backgrounds, including computer science and engineering, design, journalism and communication, medicine, mathematics, and physics.
Each participant was asked to select the most relevant chart to the anchor chart from the two candidate charts. Since some participants do not have data analysis or visualization backgrounds, we explained to them how valuable information was extracted from visualizations before the formal study. On each selection page, we provided a text introduction related to the background of the dataset to help them better understand the visualizations. After the participant completed all 30 selections, we conducted a brief interview with each participant to inquire about their decision-making criteria and collect their feedback. The entire study for each participant lasted for 30 minutes.

\subsubsection{Results}
We assessed the final result using accuracy, which measures the degree of agreement between the user's selection and the model's calculation of similar charts. The statistical analysis indicated that the average accuracy was 83.33\%, with a standard deviation of 0.073. This result suggests a high level of concordance between the model's calculations and human judgments, thus confirming the effectiveness of the Chart2Vec model in capturing contextual relationships among charts. Participants also reported that during testing they found that both candidate charts were quite relevant to the control chart and required careful analysis of the chart content. They mainly looked for similarities in two aspects: chart type and chart keywords (relevant data variables). Participants from computer science and mathematics backgrounds were more likely to examine correlation in data variables, whereas those from design backgrounds focused on correlation in terms of chart type and color scale. We took these two factors into account when designing the chart facts, and the feedback we received from users further validates the effectiveness of our design.

\subsection{Quantitative Comparison}
We conducted a quantitative comparison with two deep learning-based methods, ChartSeer and Erato, to validate the performance of Chart2Vec's chart embedding vectors in representing contextual relationships. 
The datasets and evaluation metrics utilized in this experiment were consistent with those outlined in Section~\ref{sec:ablation-study}.

\subsubsection{Model Settings}
To ensure a fair comparison with the aforementioned models, we retrained them using the same dataset as Chart2Vec. In this section, we introduce these two models respectively and provide details on the retraining process.

\textbf{ChartSeer} adopts an encoder-decoder architecture to represent data charts, taking preprocessed chart specifications as input, with data fields of Vega-Lite replaced by common markers. ChartSeer captures only the structural information from the chart specifications. To perform a fair comparison, we converted Chart2Vec's training data into ChartSeer format and retrained the model using its original framework. We utilized the encoder to represent the hidden vector of the chart. The original model achieved a reconstruction accuracy of 0.6526 with a 20-dimensional vector. To improve performance on our training data, \rv{We adjusted the configuration of ChartSeer by setting the dimension of the latent vector to 300, and kept the batch size and epochs consistent with the original, at 200 and 100, respectively. A reconstruction rate of 0.8572 was finally obtained.}

\textbf{Erato} takes chart specifications as input and converts them into sentences to obtain the initial vector representation of the chart through BERT. It then connects two fully connected layers to obtain the chart vector representation. To retrain Erato, we first converted the training data of Chart2Vec into sentence form according to the rules in Erato, and then retrained the model using its configuration with a total of 2639 steps, setting the epoch to 50 and the batch size to 64. 

\subsubsection{Results} 
As shown in Table~\ref{tab:quantitative-com}, it is evident that Chart2Vec outperforms the other two methods, ChartSeer and Erato, in terms of top-2 retrieval accuracy, top-3 retrieval accuracy, and co-occurrence values. Specifically, Chart2Vec achieves a higher top-2 retrieval accuracy, top-3 retrieval accuracy, and co-occurrence value than the means of the other two methods by 0.23, 0.25, and 0.27, respectively. These findings demonstrate that Chart2Vec is designed to effectively capture contextual associations of charts.

\rv{\rrv{We further conduct a comprehensive analysis to understand why Chart2Vec works better than the other two models.}
ChartSeer only incorporates structural information when encoding chart information, omitting specific data column names from the charts. For example, in a visualization depicting rainfall over time, it employs ``field $\rightarrow$ STR'' and ``field $\rightarrow$ NUM'' to represent rainfall and time information, respectively. In contrast, Chart2Vec not only extracts structural chart information but also takes into account the semantic information of real words. This richer representation enables Chart2Vec to explore more profound connections when capturing the contextual relationships between charts. Erato focuses on the semantic information within charts, converting the declarative language of visual charts into a sequence of strings. Additionally, it utilizes a linear interpolation loss function to map adjacent charts as a straight line in vector space. However, this approach can lead to stacking adjacent charts with similar semantics that are located in different multiple charts. Conversely, Chart2Vec supplements linear interpolation with triplet loss, bringing adjacent charts closer together while distancing them from charts located in different contexts.}

\begin{table}[!ht]
   \caption{The results of the quantitative comparison.}
\renewcommand\arraystretch{1.35}
\resizebox{\columnwidth}{!}{
\begin{tabular}{lccccc}
\hline
\multicolumn{1}{c}{Embedding Method} & \begin{tabular}[c]{@{}c@{}}Top-2 Retrieval\\ Accuracy\end{tabular} & \begin{tabular}[c]{@{}c@{}}Top-3 Retrieval\\ Accuracy\end{tabular} & Co-occurrence  & \begin{tabular}[c]{@{}c@{}} Memory\\consumption  \end{tabular}  &\begin{tabular}[c]{@{}c@{}} Time\\consumption  \end{tabular}               \\ \hline
\textbf{Chart2Vec}                            & \textbf{0.63}                                                      & \textbf{0.73}                                                      & \textbf{0.81} & 1241 MiB & 27min15s          \\
ChartSeer                            & 0.39                                       &0.47                                       &0.54 & 303 MiB & 11min15s\\
Erato                                & 0.42                                       &0.50                                       & 0.55 & 1847 MiB & 27min37s \\ \hline
\end{tabular}
}
 \label{tab:quantitative-com}
\end{table}

%% file: sections/6discussion.tex
\section{Discussion}
In this section, we discuss the limitations of the Chart2Vec model, lessons learned during the design process, and the model's generalizability and potential applications.

\subsection{Limitations and Lessons}
In this paper, we take a new perspective on the use of representation learning models to characterize visualizations in order to learn their structural, semantic, and contextual information. We conclude the limitations of Chart2Vec and summarize the key takeaways from three perspectives: the importance of context, customizing representation models, and refining them for specific visualization tasks. We also suggest potential solutions and areas for future research in these aspects.


\textbf{Necessity of Contextual Information.} 
Studies on the relationships between multi-view visualizations primarily focus on three aspects: data relationships~\cite{cantu2017identi}, spatial relationships~\cite{chen2020composition}, and interaction relationships~\cite{belo2014restruct}. In this paper, we adopt a contextual perspective to investigate the relationships between multi-view visualizations, taking into account both data and spatial relationships. This approach facilitates a more comprehensive exploration of the underlying patterns and provides deeper insights into the data. To incorporate contextual information, we curated a dataset of high-quality context-aware visualizations extracted from \rv{data stories and dashboards.} We then trained the model with shared parameters to embed the contextual information into the vector representation. As a result, in subsequent tasks such as visualization recommendation or retrieval, the fine-tuned model can leverage its learned contextual relationships to suggest or retrieve relevant visualizations. \rrv{Moreover, with the increasing interest in presenting various logic orders in exploratory visual analytics, one possible research direction is to categorize the contextual relationships and then design representation learning models based on this categorization to support more accurate downstream tasks.}


Given that the context-aware dataset we collected comprises multi-view visualizations from data stories and dashboards, contextual relationships are reflected in the proximity of spatially adjacent charts and the co-occurrence of charts within the same \rv{multi-view visualization}. At the level of data content, we regard the structural and semantic information of the visualizations as crucial information for establishing a connection between contextually related visualizations. 
Furthermore, the post-study interviews with the participants provided further support for the importance of integrating both structural and semantic information. Several participants noted that when selecting the most relevant visualizations, they not only consider the presentation of the visualizations but also the fact correlations between them. \rrv{Subsequent research can further integrate explainable AI technologies, such as feature attribution and saliency methods~\cite{wang2020saliency}, to gain insight into which part of the partial input contributes most to the final result.}

\textbf{Representation Model Customization.}
In the field of natural language processing, representation learning models have become increasingly popular by providing a way to transform textual data into vector or matrix forms, which enables computers to efficiently process and analyze data. With the prevalence of data-driven analysis and the development of automatic visualization techniques, the number of visualizations is also growing rapidly, making visualization itself an important data format~\cite{wu2021ai4vis}. Accordingly, representation learning helps transform visualizations into a data format that can be efficiently processed and analyzed using vectors. Following the footsteps of pre-trained language models, like Word2Vec~\cite{mikolov2013efficient} and BERT~\cite{devlin2018bert}, we developed a visualization representation learning model that caters to the unique features of visualization tasks and training data. 

To tailor the representation model to the needs of visualizations, we incorporated location information and constructed a task-specific training set. We utilized positional encoding to combine the structural and semantic information of the visualization. Furthermore, we adopted the triplet loss, which is commonly used in contrast learning for unsupervised learning tasks. Selecting appropriate negative samples is also crucial in this step. Since the information conveyed in visualizations heavily depends on the data from the original datasets, and different datasets tend to have diverse data distributions and attributes, it is straightforward for the model to differentiate between negative examples derived from dissimilar datasets. Therefore, when creating the training dataset, we selected negative examples from other multi-view visualizations of the same dataset or datasets of the same category. 
To preserve contextual information and obtain a more general representation, statistical information from the original dataset was not considered. However, if the downstream visualization task is highly dependent on the data statistics of the original dataset, the model architecture could be adjusted to incorporate the data schema.


\textbf{Fine-tuning for Downstream Visualization Tasks.}
The Chart2Vec model is trained using a diverse and broad training dataset. The dataset consists of 849 different data stories and 249 dashboards generated from the datasets covering 10 different domains. This approach allows Chart2Vec to learn generic multi-view visualization contexts and can be fine-tuned with small-sample datasets for specific tasks. Consequently, it can be utilized in a wide range of tasks related to context-aware visualization, leveraging state-of-the-art machine learning techniques~\cite{sun2019meta}. For example, Chart2Vec can be used to recommend relevant visualizations to users in exploratory visual analytics by concatenating sequence models such as Transformers~\cite{vaswani2017attention}. These models can be further enhanced through fine-tuning using small-sample training data derived from exploratory visual analytics sequences, which holds the potential to offer more personalized recommendations. This also presents an exciting research direction for future work.
Although we have selected training data covering 10 different topics, there is still potential for further optimization. In order to create an even more universal model, we need to increase the diversity of our training data, including dataset domains, user backgrounds, and visualization formats. Currently, the size of our training data is relatively small compared to pre-trained models in natural language processing. In the future, we plan to collaborate with well-established BI software companies to collect more high-quality multi-view visualizations created by professional data analysts, thus expanding our training data and improving the model's performance.

\subsection{Generalizability and Application Scenarios}
We discuss the generability of Chart2Vec for visualizations in other formats and some potential application scenarios.

\textbf{Visualizations in Other Formats.} The Chart2Vec model was trained using data stories from Calliope and dashboards from Tableau Public, but it can be applied to a variety of multi-view visualizations with contextual relationships, such as infographics and sequences of exploratory visualization analysis. In addition, the proposed input embedding format can be easily obtained by \rv{syntactically parsing other visualization grammar forms, such as Vega-Lite~\cite{satyanarayan2017vega} and Draco~\cite{moritz2018formalizing}}. With recent advances in information retrieval from raw images, we could also consider retraining the model using other types of visualizations by converting them into the chart fact format.

\textbf{Application Scenarios.} The Chart2Vec model is designed to transform visualizations into high-dimensional vectors, allowing for the computation of similarities \rrv{based on the contextual relevance between charts, which was previously difficult to quantify.}
This contextual computation of visualizations unlocks a wide range of practical applications, such as pattern mining, visualization recommendation, and visualization retrieval. 
First, by downsampling the visualization embeddings into a two-dimensional space, we can cluster visualizations with contextual relationships. The proximity of charts in the cluster reflects their correlations in terms of chart presentation and data content, which can be further analyzed by experts to discover their design patterns or explore data patterns encoded in the visualization ensembles.
Second, as narrative dashboards are becoming more and more popular in various areas, there is an emerging demand for automatic recommendations of context-aware visualization. For example, Medley~\cite{pan2023medley} recommends visual collections based on the user's intent, while these collections will be sorted based on the relevance of the attributes of interest to the user and the activity canvas being edited. 
\rv{In practice, Chart2Vec has also been adapted to help recommend visualization dashboards in a BI tool with over 100,000 users for the tech company.}
Moreover, during the creation of a data story or a dashboard, users may encounter the need to replace an unsatisfactory chart with a relevant one. By computing the distance between visualization embeddings, users can narrow down the search range and choose a suitable replacement.

\rrv{To illustrate how Chart2Vec works, we demonstrate a real-world use case within the context of a tech company. First, developers fine-tune the Chart2Vec model using data collected on its platform from multi-view visualizations created by users. The Chart2Vec model is integrated into the BI tool's recommendation functionality for users who want to analyze data using multi-view visualizations, enabling them to build related charts more efficiently. Users begin by selecting the dataset to be analyzed from the database. They then have the option to generate a series of single-chart visualizations using the system's internal pre-defined functions. Next, the system utilizes Chart2Vec to generate chart vectors that form the search space. As users add a single visualization to the creation panel and click the recommendation button, the system computes a similarity metric. This metric measures the contextual relevance of the chart vectors by calculating the distance between them and arranges them in order of relevance. Users can then select and add the recommended charts to the authoring panel.}

%% file: sections/7conclusion.tex
\section{Conclusion}
In this paper, we proposed Chart2Vec, a context-aware representation model that learns a universal embedding of \rrv{visualizations}, which is capable of extracting context-aware information and enabling various downstream applications such as recommendation and storytelling. We collected a context-aware visualization dataset consisting of \rv{6014 visualizations from 1098 multi-view visualizations.} Based on the four-level model of semantic content~\cite{lundgard2021accessible}, we extracted both structural and semantic information from multi-view visualizations. To better retrieve the contextual information of chart embeddings in context-aware scenarios, we adopted a multi-task training strategy that combines supervised and unsupervised learning tasks to advance the performance. We conducted a series of experiments to validate the usability and effectiveness of the model, including an ablation study, a user study, and a quantitative comparison with existing methods. In addition, we discussed the lessons learned and potential future directions. 


%% file: sections/acknowledgement.tex
\section*{ACKNOWLEDGMENTS}
Nan Cao is the corresponding author. This work was supported in part by NSFC 62372327, 62072338, NSF Shanghai 23ZR1464700, and Shanghai Education Development Foundation ``Chen-Guang Project'' 21CGA75. We would like to thank all the reviewers for their valuable feedback.